\documentclass[twocolumn,amsmath,amssymb,aps,pra,superscriptaddress,showpacs,tightenlines]{revtex4}
\usepackage{amsmath}
\usepackage{amsfonts}
\usepackage{graphicx}
\usepackage{epsfig}
\usepackage{color}
\usepackage{bm}
\usepackage[colorlinks,citecolor=blue]{hyperref}

\begin{document}

\title{Nonreciprocal high-order sidebands induced by magnon Kerr nonlinearity}

\author{Mei Wang}
\email{wangmei\_2@vip.163.com}
\affiliation{School of Electrical Engineering, Wuhan Polytechnic University, Wuhan, 430040, the People's Republic of China}
\author{Cui Kong}
\affiliation{College of Physics and Electronic Science, Hubei Normal University, Huangshi, 435002, the People's Republic of China}
\author{Zhao-Yu Sun}
\affiliation{Department of mathematics and Physics, Wuhan Polytechnic University, Wuhan, 430040, the People's Republic of China}
\author{Duo Zhang}
\affiliation{Department of mathematics and Physics, Wuhan Polytechnic University, Wuhan, 430040, the People's Republic of China}
\author{Yu-Ying Wu}
\affiliation{Department of mathematics and Physics, Wuhan Polytechnic University, Wuhan, 430040, the People's Republic of China}
\author{Li-Li Zheng}
\email{zhenglili@jhun.edu.cn}
\affiliation{Key Laboratory of Optoelectronic Chemical Materials and Devices of Ministry of Education, Jianghan University, Wuhan, 430074, People's Republic of China}

\date{\today}

\begin{abstract}
We propose an effective approach for creating robust nonreciprocity of high-order sidebands, including the first-, second- and third-order sidebands, at microwave frequencies. This approach relies on magnon Kerr nonlinearity in a cavity magnonics system composed of two microwave cavities and one yttrium iron garnet (YIG) sphere. By manipulating the driving power applied on YIG and the frequency detuning between the magnon mode in YIG and the driving field, the effective Kerr nonlinearity can be strengthened, thereby inducing strong transmission non-reciprocity. More interestingly, we find the higher the sideband order, the stronger the transmission nonreciprocity marked by the higher isolation ratio in the optimal detuning regime. Such a series of equally-spaced high-order sidebands have potential applications in frequency comb-like precision measurement, besides structuring high-performance on-chip nonreciprocal devices.
\end{abstract}

\maketitle

\section{Introduction}
Recently years, cavity magnonics system, benefited from its unique features, has shown potential application prospects in information processing. Especially, the vast spin excitation in YIG sphere called magnon, which plays a vital role in the system. Benefited from the high spin density, strong and ultra-strong magnon-microwave photon coupling ~\cite{Tobar2014,Bauer2019,Ruoso2017,Tang2014,Liu2019,Flatt2010}, observation of exceptional point and precision magnetometry~\cite{Ruoso2020,Yan2019} have been demonstrated in experiment. 
Furthermore, thanks to the high tunability, extensibility and low dissipation of the magnon, complex information interaction platforms~\cite{Nakamura2019} can be structured via establishing the interaction of magnon to an (artificial) atom~\cite{Nakamura2015}, acoustic phonons~\cite{Agarwals2019,Jaswa1998,Tang2016,Wang2017,Wang2020}, optical photons~\cite{Marquardt2016,Ferguson2015}, or superconducting coplanar microwave resonator~\cite{Goennenwein2013}.
In addition to above merits, another advantage of the system is the Kerr nonlinearity of magnon originated from the magnetocrystalline anisotropy in YIG sphere. Based on Kerr nonlinearity, a series of studies~\cite{You2018,Kons2015,Wu2019,Zhang2020} have been carried out, and one of the most notable is the nonreciprocity.

\textit{Nonreciprocity}, associated with the breaking of Lorentz reciprocity~\cite{Potton2004,Deck2018}, is manifested as asymmetric propagation of signal in two opposite directions. Nonreciprocal physics has vast and valuable applications in communication technology and information processing. For examples, one-way signal communication and manufacturing chip-scale nonreciprocal devices: isolator, circulator, diode. Stimulated by the bright prospect,
considerable efforts has been made in the study of nonreciprocal physics. One of the most watched is the nonreciprocity of electromagnetic fields~\cite{Caloz2018, Asadchy2020}, which is mainly concerned with the nonreciprocal propagation of electromagnetic fields in the microwave and optical regions.
Microwave and optical nonreciprocity are both typically achieved from the magneto-optic crystal~\cite{H2000,Ram2007,Raghu2008,Osgood2008,Solja2009} relying on Faraday-rotation~\cite{Fan2003, Assanto2008, Sai2013} to break the Lorentz reciprocity. However, the tricky thing is the bulky and heavy structure of the magneto-optic crystal, that poses a serious hurdle to flexibly control and chip-scale integration. Over the past few decades, alternative mechanisms towards realizing nonreciprocal propagation at microwave and optical frequencies have been proposed. The most promising ones are based on nonlinearity~\cite{Review1,Review2,Review3,Review4,Review5,Review6,Review7,Aumentado2017,Ustinov2012,Kippen2018},  reservoir engineering~\cite{Clerk2017,Clerk2015}, parametric time modulation~\cite{Alu2017,Aumentado2014}.

Inspired by previous studies, based on nonlinearity mechanism, we propose a flexible scheme to achieve robust nonreciprocity of the first-, second-, and third-order sidebands at microwave frequencies. This approach is implemented in a three-mode cavity magnonics system, which consists of two coupled microwave cavities and a YIG (or another suitable magnonic material) sphere. Due to the magnon Kerr nonlinearity in YIG sphere, strong nonreciprocity of high-order sidebands can be obtained by flexibly manipulating the driving power injected on YIG sphere and the frequency detuning between magnon mode (in YIG) and the driving field. On one hand, the driving power applied on YIG affects excitation of magnon that determines the effective strength of the Kerr nonlinearity. It therefor induces varying degrees asymmetric responses of the system in two opposite directions.
On the other hand, frequency detuning between the driving field and the magnon mode also affects magnon excitation in YIG sphere. Thus the effective Kerr nonlinearity will be influenced, as well as degree of the asymmetric responses in two directions.
More interestingly, in the optimal parameters regime, the greater the order of the sideband, the stronger transmission nonreciprocity marked by the enhanced isolation ratio.
Such a series of equally-spaced high-order sidebands can be applied in frequency comb-like precision measurement, Besides structuring nonreciprocal devices.
In addition, frequencies of the sidebands avoid the main cavity frequency and this will strongly inhibit dissipation. The system parameters are referenced from recent experiments~\cite{You2018,Wang2016}, specially, Kerr nonlinear coefficient ($K$ reaches $10^2$\,nHz refereing to~\cite{Zhang2019}) in the system is highly dependent on the specific experimental implementation. This study broadens applications of the cavity magnonics system in information processing.

This paper is organized as follows: In Sec. II, we introduce theoretical scheme of the three-mode cavity magnonics system and its effective Hamiltonian with external pump fields. Then we derive transmission coefficients and isolation ratios of the first-, second- and third-order sidebands in two cases. In Sec. III, we study transmission nonreciprocity of the high-order sidebands through manipulating the driving field on the YIG sphere and the frequency detuning between driving field and the magnon mode in YIG sphere. Conclusions are finally drawn in Sec. IV.
\begin{figure}[htb]
\centering\includegraphics[width=8.4cm]{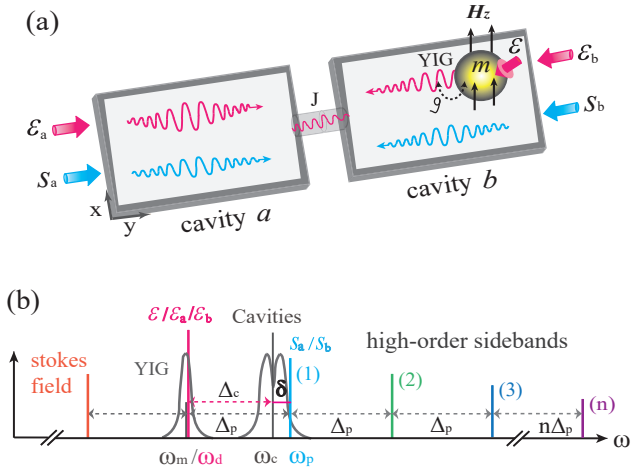}
\caption{(Color online) (a) Schematic diagram of a three-mode cavity magnonics
system consisting of two coupled microwave cavities and a YIG sphere. (b)
Frequency spectrogram of the cavity magnonics system. Here the driving field (with amplitude $\varepsilon$) applied on YIG sphere and the control fields (with amplitudes $\varepsilon_a$, $\varepsilon_b$) injected from microwave cavities have the same frequency $\omega_d$.
Frequencies of the control fields and the probe fields (with amplitudes $s_a$, $s_b$) are respectively detuned from the cavities' frequency by $\Delta_c$ and $\delta$. And the frequency detuning between the probe field and the control field is $\Delta_p$. There are higher-order sidebands at frequency $\omega_d+n\Delta_p$, where $n$ is an integer representing the sideband order.}
\label{fig:1}
\end{figure}

\section{The Model and formula derivation}
We consider a three-mode cavity magnonics system~\cite{You2018} as shown in Fig.~\ref{fig:1}(a). It is composed of two three-dimensional rectangular microwave cavities ($a$ and $b$, frequency $\omega_c$) and one YIG sphere, where collective spin excitation is quantified as magnon ($m$, frequency $\omega_m$).
The cavities are linearly coupled to each other with coupling rate $J$; and each of them is driven and detected by a bichromatic microwave source, i.e., a strong control field with amplitude $\varepsilon_a$ ($\varepsilon_b$) at frequency $\omega_d$ and a weak probe field $s_a$ ($s_b$) at frequency $\omega_p$.
The YIG sphere, located in one cavity, is exposed to a uniformly DC magnetic bias field (${ H}_z$) in the $z$ direction. Meanwhile, it interacts with one cavity with strength $g$, and is pumped by a microwave driving field with amplitude $\varepsilon$ and frequency $\omega_d$. Note that amplitudes of the input microwave fields can be converted into corresponding powers, i.e., $P_{j}=\hbar\omega_d{\varepsilon}^2_j$, $P=\hbar\omega_d{\varepsilon}^2$ and $P_{s_j}=\hbar\omega_p{s}^2_j$ ($j=a,b$) with control power $P_{j}$, driving power $P$ and probe power $P_{s_j}$.

In the rotating frame, including all external microwave sources, the system Hamiltonian (refering to~\cite{You2018,Wang2016}) can be written as
\begin{eqnarray}
\hat{H}_{\rm tot}/\hbar&=&\Delta_{c}(\hat{a}^{\dagger}\hat{a}+\hat{b}^{\dagger}\hat{b})+\Delta_{m}\hat{m}^{\dagger}\hat{m}+K\hat{m}^{\dagger}\hat{m}\hat{m}^{\dagger}\hat{m}
\nonumber\\&&
+J(\hat{a}^{\dagger}\hat{b}+\hat{a}\hat{b}^{\dagger})+g(\hat{b}^{\dagger}\hat{m}+\hat{b}\hat{m}^{\dagger})
\nonumber\\&&
+i\sqrt{\eta_{a}\kappa_{a}}\varepsilon_{a}(\hat{a}^{\dagger}-\hat{a})+i\sqrt{\eta_{b}\kappa_{b}}\varepsilon_{b}(\hat{b}^{\dagger}-\hat{b})
\nonumber\\&&
+i\sqrt{\eta_{a}\kappa_{a}}s_{a}(\hat{a}^{\dagger}e^{-i\Delta_p t}-\hat{a}e^{i\Delta_p t})
\nonumber\\&&
+i\sqrt{\eta_{b}\kappa_{b}}s_{b}(\hat{b}^{\dagger}e^{-i\Delta_p t}-\hat{b}e^{i\Delta_p t})
\nonumber\\&&
+i\sqrt{\eta_{m}\kappa_{m}}\varepsilon(\hat{m}^{\dagger}-\hat{m}),\label{e1}
\end{eqnarray}
here $\Delta_{m}=\omega_{m}-\omega_d$, $\Delta_p=\omega_p-\omega_c$ and $\Delta_{c}=\omega_{c}-\omega_d$ is constant. In addition, the frequency detunings satisfy relationship $\Delta_p=\Delta_c+\delta$, where $\delta=\omega_p-\omega_c$ refereing to Fig.\,\ref{fig:1}(b). The third term in Hamiltonian is magnon Kerr nonlinear term originating from the magnetocrystalline anisotropy in the YIG sphere. The Kerr nonlinear coefficient $K=\mu_0 K_{an}{\gamma}^2/(M^2V_m)$, where $\mu_0$ is the magnetic permeability of free space, $K_{an}$ is the first-order anisotropy constant, $\gamma$ is the gyromagnetic ratio, $M$ is the saturation magnetization, and $V_m$ is the volume of the YIG sphere.
$\kappa_j$ ($j=a,b,m$) represents the total dissipation of any microwave cavity or magnon mode with cavity (magnon) coupling parameter $\eta_j$.

In this work, we employ analytic solution method to solve Heisenberg-Langevin equations. It not only describes the dynamical evolution of the cavity magnonics system, but also responses the expectation values of all operators, namely, $\langle \hat{o}\rangle=o$ ($\hat{o}$ is a normal annihilation operator). Taking account damping progresses in the system, Heisenberg-Langevin equations of the system can be expressed as
\begin{subequations}
\label{e2}
\begin{eqnarray}
\dot{{a}}&=&-(i\Delta_{c}+\frac{\kappa_{a}}{2}){a}-iJ{b}+\sqrt{\eta_{a}\kappa_{a}}\varepsilon_{a} \nonumber\\
&&+\sqrt{\eta_{a}\kappa_{a}}s_{a}e^{-i\Delta_p t},\label{e2a}
\\
\dot{{b}}&=&-(i\Delta_{c}+\frac{\kappa_{b}}{2}){b}-iJ{a}-ig{m}+\sqrt{\eta_{b}\kappa_{b}}\varepsilon_{b} \nonumber\\
&&+\sqrt{\eta_{b}\kappa_{b}}s_{b}e^{-i\Delta_p t},\label{e2b}
\\
\dot{{m}}&=&-(i\Delta_{m}+\frac{\kappa_{m}}{2}){m}-ig{b}-i(2K|{m}|^2+K){m} \nonumber\\
&&+\sqrt{\eta_{m}\kappa_{m}}\varepsilon.\label{e2c}
\end{eqnarray}
\end{subequations}
Equations ~(\ref{e2a}) and (\ref{e2b}) describe the dynamics of cavities $a$ and $b$. Equation ~(\ref{e2c}) describes the dynamics of the magnon mode.
The cubic term $-2iK|{m}|^2{m}$ in Equation~(\ref{e2c}) presents the nonlinearity of the system. It is derived from the Kerr nonlinearity term in Hamiltonian~(\ref{e1}) with factorizing averages approximation, i.e., $\langle bc\rangle=\langle b\rangle\langle c\rangle$. The input quantum and thermal noise terms are ignored here because their mean values are zero.

For existence of the nonlinear term, equations~(\ref{e2a})-(\ref{e2c}) are inherently nonlinear and cannot be solved directly and precisely. However, due to
the strength of the control fields ($5\,\,m$W) are far stronger than the probe fields ($1\,\,n$W) in the system, we can take perturbation method
\begin{figure}[htb]
\centerline{\includegraphics[width=8.4cm]{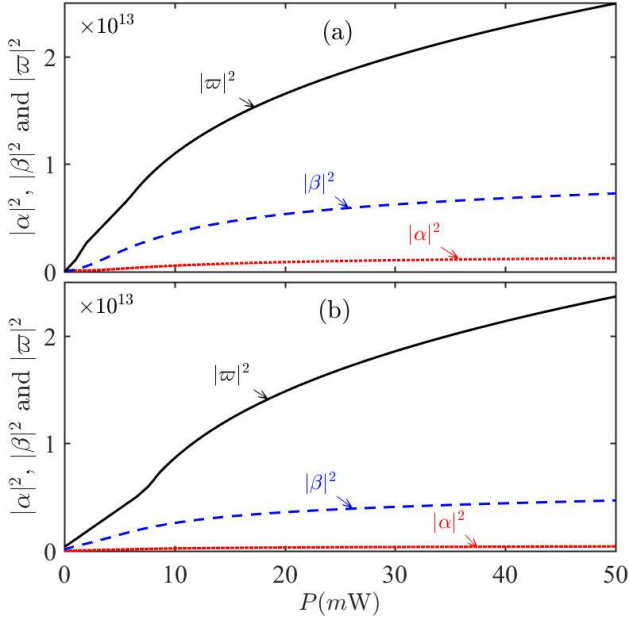}}
\caption{(Color online) The steady-state microwave photon numbers $|\alpha|^2$, $|\beta|^2$ and magnon number $|\varpi|^2$ versus with the resonant driving power $P$ on the magnon mode in two cases. (a) case 1: $P_{a}=5\,\, m$W, $P_{s_a}=1\,\,n$W, $P_{b}=P_{s_b}=0$; (b) case $2$: $P_{b}=5\,\, m$W, $P_{s_b}=1\,\,n$W, $P_{a}=P_{s_a}=0$. Other parameters are given as $\omega_c/2\pi=10.1$\,\,GHz, $\omega_m/2\pi=\omega_d/2\pi=10.06$\,\,GHz, $\Delta_c/2\pi=40$\,\,MHz, $\Delta_m=0$, $\kappa_a/2\pi=3.8$\,\,MHz, $\kappa_b/2\pi=5$\,\,MHz, $\kappa_m/2\pi=20$\,\,MHz, $\delta/2\pi=-12$\,\,MHz, $K/2\pi=3.9\times10^{-7}$\,\,Hz, $J/2\pi=12$\,\,MHz, $g/2\pi=8$\,\,MHz.}
\label{fig:2}
\end{figure}
to solve equations~(\ref{e2}). That is dividing each operator into two parts: a steady-state mean value and a perturbation term, i.e., $a ={\alpha}+\delta a$, $b = {\beta}+\delta b$, $m={\varpi}+\delta m$. It means that the steady-state mean values of the operators are determined by the strong control fields, and the perturbation terms are caused by the weak probe fields. When the probe fields are absent, with $d{\alpha}/dt=0$, $d{\beta}/dt=0$ and $d{\varpi}/dt=0$, the mean values of the operators from the steady-state equations can be derived as
\begin{eqnarray}
{\alpha}&=&\frac{\sqrt{\eta_{a}\kappa_{a}}\varepsilon_{a}-iJ{\beta}}{i\Delta_{c}+\frac{\kappa_{a}}{2}},
\\
{\beta}&=&\frac{g(\Delta_{c}-i\frac{\kappa_{a}}{2}){\varpi}-iJ\sqrt{\eta_{a}\kappa_{a}}\varepsilon_{a}+(i\Delta_{c}+\frac{\kappa_{a}}{2})\sqrt{\eta_{b}\kappa_{b}}\varepsilon_{b}}{(i\Delta_{c}+\frac{\kappa_{a}}{2})(i\Delta_{c}+\frac{\kappa_{b}}{2})+J^{2}},
\nonumber\\&&
\\
{\varpi}&=&\frac{-gJ\sqrt{\eta_{a}\kappa_{a}}\varepsilon_{a}+g(\Delta_{c}-i\frac{\kappa_{a}}{2})\sqrt{\eta_{b}\kappa_{b}}\varepsilon_{b}}{\left[(i\Delta_{c}+\frac{\kappa_{a}}{2})(i\Delta_{c}+\frac{\kappa_{b}}{2})+J^{2}\right]{\Theta}}
+\frac{\sqrt{\eta_{m}\kappa_{m}}\varepsilon}{{\Theta}},
\nonumber\\&&
\label{e3}
\end{eqnarray}
where we have defined
\begin{eqnarray}
\label{e5}
\Theta&=&\frac{g^{2}(i\Delta_{c}+\frac{\kappa_{a}}{2})}{(i\Delta_{c}+\frac{\kappa_{a}}{2})(i\Delta_{c}+\frac{\kappa_{b}}{2})+J^{2}}
+\frac{\kappa_{m}}{2}
\nonumber\\&&
+i(\Delta_{m}+2K|\varpi|^{2}+K).
\end{eqnarray}

In Fig.~\ref{fig:2}, evolutions of the steady-state microwave-photon numbers and the magnon number versus the resonant driving power on magnon mode are shown in two different cases. Figure~\ref{fig:2}(a) corresponds to case $1$, only cavity $a$ is driven by the bichromatic microwave fields and cavity $b$ is free of driven. In case $2$, the same bichromatic microwave fields are only applied on cavity $b$ in Fig.~\ref{fig:2}(b). As we can see the steady-state microwave photon numbers and the magnon number in case $1$ are larger than those in case $2$ even with the same driving power $P$. In addition, in both cases the microwave-photon numbers in cavity $a$ are smaller than that in cavity $b$, though the dissipation of cavity $a$ is lower than that of cavity $b$ and the magnon mode, i.e., $\kappa_a<\kappa_b,\kappa_m$ (on the basis of $\frac{g}{J}>\frac{1}{3}$ and $K\geq1 n$Hz). The result illustrates that the counterintuitive asymmetric responds of the steady-state mean values are affected by the nonlinearity term in equations~(\ref{e2}). Meanwhile, it also plays an important role in evolutions of the perturbation terms.

Expanding equations~(\ref{e2a})-(\ref{e2c}) with the ansatz of operators and removing the steady-state terms, the perturbation terms can be expressed by the linearized Heisenberg-Langevin equations
\begin{subequations}
\begin{eqnarray}
\delta \dot{a}&=&-(i\Delta_{c}+\frac{\kappa_{a}}{2})\delta a-iJ\delta b+\sqrt{\eta_{a}\kappa_{a}}s_{a}e^{-i\Delta_p t},\label{e7a}
\\
\delta \dot{b}&=&-(i\Delta_{c}+\frac{\kappa_{b}}{2})\delta b-iJ\delta a-ig\delta m
\nonumber\\&&
+\sqrt{\eta_{b}\kappa_{b}}s_{b}e^{-i\Delta_p t},\label{e7b}
\\
\delta\dot{m}&=&-[i(\Delta_{m}+4K|{ \varpi}|^2+K)+\frac{\kappa_{m}}{2}]\delta m-ig\delta b
\nonumber\\&&
-i2K({\varpi}^{2}\delta m^{*}+{\varpi}^{*}\delta m^{2}+2{\varpi}\delta m^{*}\delta m )
\nonumber\\&&
-i2K\delta m^{*}\delta m^{2},
\label{e7c}
\end{eqnarray}
\label{e7}
\end{subequations}
the quadratic terms $-i2K{\varpi}^{*}\delta m^{2}$, $-i4K{\varpi}\delta m^{*}\delta m$ in equation~(\ref{e7c}) denote the system nonlinearity and can not be sneezed at. They inevitably cause asymmetric responses of the perturbation terms, just as the responses of the steady-state mean values in Fig.~\ref{fig:2}.
However, the cubic term $-i2K\delta m^{*}\delta m^{2}$ in equations~(\ref{e7}) is so small that it will be ignored in the following calculation.

To study asymmetric responses of the perturbations from a new perspective, we explore the nonreciprocal transmission of the output fields, mainly including the first-, second- and third-order sidebands, in case $1$ and case $2$.
To get analytical solutions of these high-order sidebands, we define the perturbation terms have the following forms
\begin{subequations}
\begin{eqnarray}
\delta a&=&{A^{-}_{1}}e^{-i\Delta_p t}+{A^{+}_{1}}e^{i\Delta_p t}+{A^{-}_{2}}e^{-2i\Delta_p t}+{A^{+}_{2}}e^{2i\Delta_p t}
\nonumber\\&&
+{A^{-}_{3}}e^{-3i\Delta_p t}+{A^{+}_{3}}e^{3i\Delta_p t}, \label{e8a}
\\
\delta b&=&{B^{-}_{1}}e^{-i\Delta_p t}+{ B^{+}_{1}}e^{i\Delta_p t}+{B^{-}_{2}}e^{-2i\Delta_p t}+{ B^{+}_{2}}e^{2i\Delta_p t}
\nonumber\\&&
+{B^{-}_{3}}e^{-3i\Delta_p t}+{ B^{+}_{3}}e^{3i\Delta_p t},\label{e8b}
\\
\delta m&=&{ M^{-}_{1}}e^{-i\Delta_p t}+{M^{+}_{1}}e^{i\Delta_p t}+{ M^{-}_{2}}e^{-2i\Delta_p t}+{M^{+}_{2}}e^{2i\Delta_p t}
\nonumber\\&&
+{ M^{-}_{3}}e^{-3i\Delta_p t}+{M^{+}_{3}}e^{3i\Delta_p t},\label{e8c}
\end{eqnarray}
\label{e8}
\end{subequations}
this ansatz is rooted from the internal mechanism of the system, where a series of  high-order sidebands~\cite{Xiong2013} with frequency $\omega_d+n\Delta_p$ ($n=1,2,3...$) are generated. $n$ represents the sideband order as shown in Fig.~\ref{fig:1}(b). The negative and positive exponents of the powers are in pairs and respectively denote the upper and lower sidebands. For examples, the term ${A^{-}_{1}}e^{-i\Delta_p t}$ denotes the first-order upper sideband with coefficient $A^{-}_{1}$, and ${A^{+}_{1}}e^{i\Delta_p t}$ corresponds to the first-order lower sideband (the Stokes field~\cite{Wu2020} in Fig.~\ref{fig:1}(b)) with coefficient $A^{+}_{1}$. For the sake of simplicity, in what follows we only take the high-order upper sidebands into consideration, because of similar characteristics in the lower sidebands. Of course the corresponding lower sidebands can also be studied by utilizing the same method.

With the ansatz we can analytically solve equations~(\ref{e7}), the coefficients of the high-order sidebands can be derived, respectively. Firstly, the coefficients of the first-order upper sideband are derived as
\begin{subequations}
\begin{eqnarray}
B_{1}^{-}&=&\frac{i\sqrt{\eta_{b}\kappa_{b}}s_{b}(\Delta_p-\Delta_{c}+i\frac{\kappa_{a}}{2})(i\mathcal{C}-\Delta_p+2\varpi^{2}K\mathcal{L})}{g^{2}(\Delta_p-\Delta_{c}+i\frac{\kappa_{a}}{2})+(i\mathcal{C}-\Delta_p+2{\varpi}^{2}K\mathcal{L})\mathcal{S}_{(-)}}
\nonumber\\&&
+\frac{iJ\sqrt{\eta_{a}\kappa_{a}}s_{a}(i\mathcal{C}-\Delta_p+2\varpi^{2}K\mathcal{L})}{g^{2}(\Delta_p-\Delta_{c}+i\frac{\kappa_{a}}{2})+(i\mathcal{C}-\Delta_p+2{\varpi}^{2}K\mathcal{L})\mathcal{S}_{(-)}},\label{e9a}
\nonumber\\&&
\\
A_{1}^{-}&=&\frac{JB_{1}^{-}+i\sqrt{\eta_{a}\kappa_{a}}s_{a}}{\Delta_p-\Delta_{c}+i\frac{\kappa_{a}}{2}},\label{e9b}
\\
M_{1}^{-}&=&\frac{-gB_{1}^{-}}{i\mathcal{C}-\Delta_p+2K{\varpi}^{2}\mathcal{L}},\label{e9c}
\\
M_{1}^{+}&=&\mathcal{L}^*M_{1}^{-*},\label{e9d}
\end{eqnarray}
\label{e9}
\end{subequations}
with
\begin{eqnarray}
\mathcal{C}&=&-i(\Delta_{m}+K+4K|{\varpi}|^2)-\frac{\kappa_{m}}{2},\label{e10}
\\
\mathcal{S}_{(\pm)}&=&(\Delta_p\pm\Delta_{c}+i\frac{\kappa_{a}}{2})(\Delta_p\pm\Delta_{c}+i\frac{\kappa_{b}}{2})-J^{2},\label{e11}
\\
\mathcal{L}&=&\frac{2K\mathcal{S}_{(+)}{\varpi}^{2*}}{g^{2}(\Delta_p+\Delta_{c}+i\frac{\kappa_{a}}{2})-(\Delta_p-i\mathcal{C})\mathcal{S}_{(+)}}. \label{e12}
\end{eqnarray}
For the second-order upper sideband, the coefficients are given as
\begin{subequations}
\begin{eqnarray}
B_{2}^{-}&=&\frac{g(2\Delta_p-\Delta_{c}+i\frac{\kappa_{a}}{2})M_{2}^{-}}{G},\label{e13a}
\\
A_{2}^{-}&=&\frac{JB_{2}^{-}}{2\Delta_p-\Delta_{c}+i\frac{\kappa_{a}}{2}},\label{e13b}
\\
M_{2}^{-}&=&\frac{-2KG(\varpi^{2}E^{*}+\varpi^{*}M_{1}^{-2}+2\varpi M_{1}^{-}M_{1}^{+*})}{(i\mathcal{C}-2\Delta_p+2K\varpi^{2}D^{*})G+g^{2}(2\Delta_p-\Delta_{c}+i\frac{\kappa_{a}}{2})},\label{e13c}
\nonumber\\&&
\\
M_{2}^{+}&=&DM_{2}^{-*}+E,\label{e13d}
\end{eqnarray}
\label{e13}
\end{subequations}
where
\begin{eqnarray}
D&=&\frac{2\varpi^{2}KF}{g^{2}(2\Delta_p+\Delta_{c}-i\frac{\kappa_{a}}{2})-F(2\Delta_p+i\mathcal{C})},\label{e14}
\\
E&=&\frac{2\varpi^{*}KFM_{1}^{+2}+4\varpi K FM_{1}^{+}M_{1}^{-*}}{g^{2}(2\Delta_p+\Delta_{c}-i\frac{\kappa_{a}}{2})-F(2\Delta_p+i\mathcal{C})},\label{e15}
\\
F&=&(2\Delta_p+\Delta_{c}-i\frac{\kappa_{b}}{2})(2\Delta_p+\Delta_{c}-i\frac{\kappa_{a}}{2})-J^{2},\label{e16}
\\
G&=&(2\Delta_p-\Delta_{c}+i\frac{\kappa_{b}}{2})(2\Delta_p-\Delta_{c}+i\frac{\kappa_{a}}{2})-J^{2}.\label{e17}
\end{eqnarray}
To the third-order upper sideband, the coefficients have the following forms
\begin{subequations}
\begin{eqnarray}
B_{3}^{-}&=&\frac{g(3\Delta_p-\Delta_{c}+i\frac{\kappa_{a}}{2})M_{3}^{-}}{(3\Delta_p-\Delta_{c}+i\frac{\kappa_{a}}{2})(3\Delta_p-\Delta_{c}+i\frac{\kappa_{b}}{2})-J^{2}},\label{e18a}
\\
A_{3}^{-}&=&\frac{JB_{3}^{-}}{3\Delta_p-\Delta_{c}+i\frac{\kappa_{a}}{2}},\label{e18b}
\\
M_{3}^{-}&=&\frac{2K\varpi^{2}H_{+}^{*}+H_{-}(3\Delta_c+gT^{*}-i\mathcal{C}^{*})}{(3i\Delta_p+\mathcal{C}-igQ)(-3\Delta_p-gT^{*}+i\mathcal{C}^{*})-4iK^{2}|\varpi|^{4}},
\label{e18c}\nonumber\\&&
\\
M_{3}^{+}&=&\frac{2\varpi^{2}KM_{3}^{-*}+iH_{+}}{-3\Delta_p-gB_{3}-i\mathcal{C}},\label{e18d}
\end{eqnarray}
\label{e18}
\end{subequations}
in which
\begin{eqnarray}
Q&=&\frac{g(3\Delta_p-\Delta_{c}+i\frac{\kappa_{a}}{2})}{(3\Delta_p-\Delta_{c}+i\frac{\kappa_{a}}{2})(3\Delta_p-\Delta_{c}+i\frac{\kappa_{b}}{2})-J^{2}},
\\
T&=&\frac{-g(3\Delta_p+\Delta_{c}-i\frac{\kappa_{a}}{2})}{(3\Delta_p+\Delta_{c}-i\frac{\kappa_{a}}{2})(3\Delta_p+\Delta_{c}-i\frac{\kappa_{b}}{2})-J^{2}},
\\
H_{+}&=&-4i\varpi^{*}KM_{1}^{+}M_{2}^{+}-4i\varpi K(M_{1}^{+}M_{2}^{-*}+M_{2}^{+}M_{1}^{-*}),
\nonumber\\&&
\\
H_{-}&=&-4i\varpi^{*}KM_{1}^{-}M_{2}^{-}-4i\varpi K(M_{2}^{-}M_{1}^{+*}+M_{1}^{-}M_{2}^{+*}).
\nonumber\\&&
\end{eqnarray}
It can be seen that the coefficients $A_{1}^{-}$, $B_{1}^{-}$, $A_{2}^{-}$, $B_{2}^{-}$, $A_{3}^{-}$ and $B_{3}^{-}$ of the high-order sidebands are closely dependent on the magnon mode including the steady-state amplitude and the perturbation terms.

With above analytical solutions, we calculate the output fields of each sideband in two cases. In case $1$, the control and probe fields are injected from cavity $a$ and the output fields from cavity $b$ are explored. With input-output relation, the output fields of high-order sidebands from cavity $b$ can be expressed as
\begin{eqnarray}
s_{{aout}}&=&\sqrt{\eta_{b}\kappa_{b}}B_{1}^{-}e^{-i(\omega_{d}+\Delta_p)t}+\sqrt{\eta_{b}\kappa_{b}}B_{2}^{-}e^{-i(\omega_{d}+2\Delta_p)t}
\nonumber\\&&
+\sqrt{\eta_{b}\kappa_{b}}B_{3}^{-}e^{-i(\omega_{d}+3\Delta_p)t}.\label{e19}
\end{eqnarray}
In the direction from cavity $a$ to $b$, the transmission coefficients of the first-, second- and third-order sidebands are respectively given as
\begin{subequations}
\begin{eqnarray}
t_{ba}&=&\left|\frac{\sqrt{\eta_{b}\kappa_{b}}B_{1}^{-}}{s_{a}}\right|,\label{e20a}
\\
\tau_{ba}&=&\left|\frac{\sqrt{\eta_{b}\kappa_{b}}B_{2}^{-}}{s_{a}}\right|,\label{e20b}
\\
\ell_{ba}&=&\left|\frac{\sqrt{\eta_{b}\kappa_{b}}B_{3}^{-}}{s_{a}}\right|.\label{e20c}
\end{eqnarray}
\label{e20}
\end{subequations}

In case $2$, the same control and probe fields are applied on cavity $b$ and the output fields from cavity $a$ are researched. In order to avoid confusion, we use $\mathcal{A}^{-}_j$, $\mathcal{B}^{-}_j$ and $\mathcal{M}^{-}_j$ ($j=1,2,3$) to distinguish from the coefficients in case $1$. Combining this condition and the input-output relation, output fields of the high-order upper sidebands are given as follows
\begin{eqnarray}
s_{{bout}}&=&\sqrt{\eta_{a}\kappa_{a}}\mathcal{A}_{1}^{-}e^{-i(\omega_{d}+\Delta_p)t}+\sqrt{\eta_{a}\kappa_{a}}\mathcal{A}_{2}^{-}e^{-i(\omega_{d}+2\Delta_p)t}
\nonumber\\&&
+\sqrt{\eta_{a}\kappa_{a}}\mathcal{A}_{3}^{-}e^{-i(\omega_{d}+3\Delta_p)t}.\label{e21}
\end{eqnarray}
Similarly, along the direction from cavity $b$ to $a$, the transmission coefficients of the first-, second- and third-order sidebands are defined as
\begin{subequations}
\begin{eqnarray}
t_{ab}&=&\left|\frac{\sqrt{\eta_{a}\kappa_{a}}\mathcal{A}_{1}^{-}}{s_{b}}\right|,\label{e22a}
\\
\tau_{ab}&=&\left|\frac{\sqrt{\eta_{a}\kappa_{a}}\mathcal{A}_{2}^{-}}{s_{b}}\right|,\label{e22b}
\\
\ell_{ab}&=&\left|\frac{\sqrt{\eta_{a}\kappa_{a}}\mathcal{A}_{3}^{-}}{s_{b}}\right|.\label{e22c}
\end{eqnarray}
\label{e22}
\end{subequations}

In order to more clearly describe nonreciprocal transmissions of the high-order sidebands in two cases, we introduce definition of the nonreciprocal isolation ratio, which is expressed as
\begin{eqnarray}
\label{e23}
I_o=\left|{\rm lo}g_{10}\frac{|\sqrt{\eta_{b}\kappa_{b}}B^{-}_{j}|^2}{|\sqrt{\eta_{a}\kappa_{a}}\mathcal{A}^{-}_{j}|^2}\right|,
\end{eqnarray}
here $o$ stands for either $t$, $\tau$ or $\ell$ corresponding to $j=1,2,3$. $I_t$, $I_\tau$ and $I_\ell$ respectively denote isolation ratio of the first-, second- and third-order sidebands. Then we use isolation ratio $I_o$ to quantitatively describe degree of the nonreciprocal transmissions in two cases. It is decided by the value of $I_o$: if $I_o=0$, equivalently, $|\sqrt{\eta_{b}\kappa_{b}}B^{-}_{j}|^2/|\sqrt{\eta_{a}\kappa_{a}}\mathcal{A}^{-}_{j}|^2=1$, it means transmissions of the high-order sidebands are reciprocal (symmetry) along two opposite directions. For other situations, the higher the value of $I_o$, the stronger is the nonreciprocity of the transmissions in two cases. For example, $I_o \geq1$, which illustrates transmission coefficient in one direction is at least $10$ times that of the other direction. It also shows that transmittances along two opposite directions have a strong asymmetry. This situation can be approximated as one-way transmission of high-order sidebands.

\section{Nonreciprocal transmissions of the high-order sidebands in a cavity magnonics system}
Since the transmission coefficients in equations~(\ref{e20}) and (\ref{e22}) and the isolation ratio in equation~(\ref{e23}) are mainly dominated by coefficients of each sideband, which are closely associated with the steady-state amplitude and the perturbation terms of magnon mode. Under this situation, we analyze influence of the driving power, applied on the magnon mode, and the frequency detuning between the driving field and the magnon mode on the nonreciprocal transmission of each sideband.
\begin{figure}[tbh]
\centering
\includegraphics[width=8cm]{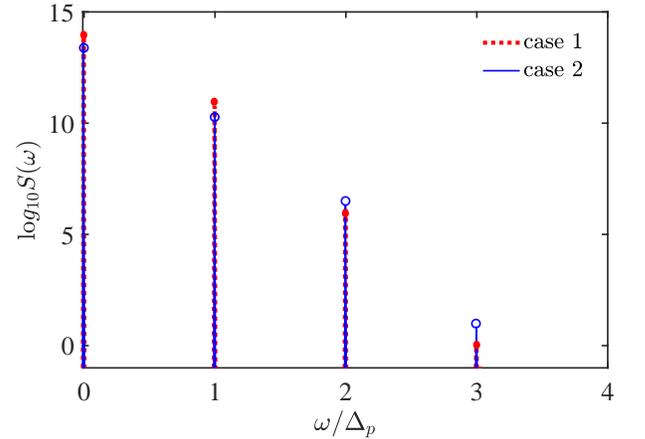}
\caption{(Color online) The logarithmic amplitudes $ {\rm lo}g_{10}{\rm S}(\omega)$ of the transmitted high-order sidebands vs frequency $\omega/\Delta_p$ in case $1$ and case $2$. Here $P=15\,\,m$W, $\Delta_p/\Delta_c=0.7$, other parameters are same as in Fig.~\ref{fig:2}.}
\label{fig:3}
\end{figure}

For the same purpose, but utilizing a different method, we first exhibit an overview comparison of the high-order sideband spectrums between case $1$ and case $2$. Here the magnon mode is resonantly driven by a $15\,\,m$W driving field. As shown in Fig.~\ref{fig:3}, spectrums of the adjacent sidebands are equally spaced apart from each other at frequency $\Delta_p$ in the rotating frame. Where $\omega/\Delta_p=0$
denotes the control field, $\omega/\Delta_p=n$ ($n=1,2,3$) corresponds to the nth-order sideband. It can be seen that there exists amplitude gap between the sidebands with the same order in case $1$ and case $2$. Scilicet, the transmissions of the high-order sidebands are nonreciprocal in two cases. What's more, amplitudes of the high-order sidebands decrease rapidly as the sideband order increases. Specifically, $S(\Delta_p)/S(0)<10^{-3}$, $S(2\Delta_p)/S(0)<10^{-7}$ and $S(3\Delta_p)/S(0)<10^{-12}$
in both cases. This proves that it is valid to regard high-order sidebands as perturbation
compared with the strong control field. Then, we analyze in detail the transmission nonreciprocity of each sideband in the following content.

\subsection{Dependence of nonreciprocal transmissions for high-order sidebands on the resonant driving power $P$}
\begin{figure}[tbh]
\centering
\includegraphics[width=8.5cm]{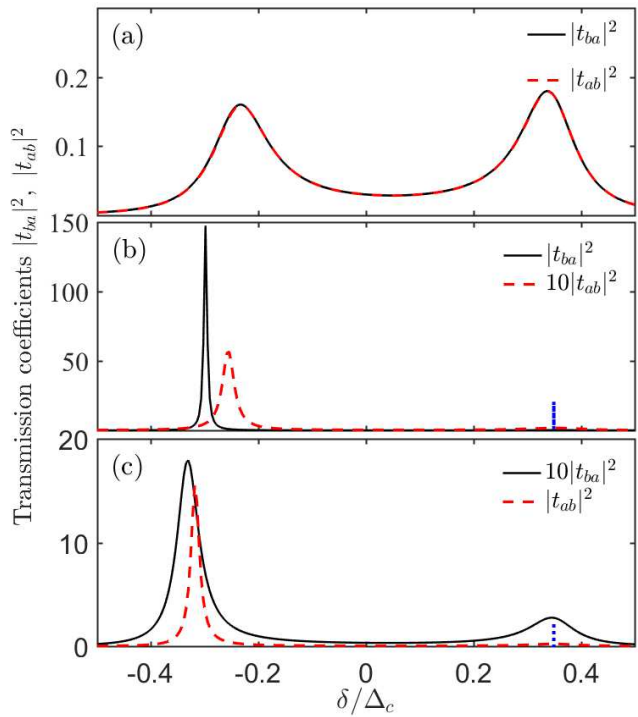}
\caption{(Color online) Transmission coefficients $|t_{ba}|^2$ and $|t_{ab}|^2$ of the first-order sideband vs frequency detuning $\delta/\Delta_c$. To observe transmission coefficients clearly, $|t_{ab}|^2$ in (b) and $|t_{ba}|^2$ in (c) have been increased by one order of magnitude. The driving powers $P=0$, $20\,\,m$W, $30\,\,m$W in (a), (b) and (c), respectively. Other parameters are same as that in Fig.~\ref{fig:2}.}
\label{fig:4}
\end{figure}

Figure~\ref{fig:4} shows transmission coefficients $|t_{ba}|^2$ and $|t_{ab}|^2$ vary with frequency detuning $\delta/\Delta_c$ under different resonant driving powers $P$. In Fig.~\ref{fig:4}(a), we choose driving power $P=0$. Under this situation, $|t_{ba}|^2$ and $|t_{ab}|^2$ are smaller than $0.2$ and evolutions of them are synchronous.
It means the transmissions of the first-order sideband in two cases are reciprocal when the driving field on magnon mode is absent. Increasing driving power to $20\,\,m$W in Fig.~\ref{fig:4}(b), intuitively, only a peak with value $150$ of $|t_{ba}|^2$ is seen at $\delta/\Delta_c=-0.3$; and a peak value $5.6$ of $|t_{ab}|^2$ is shown at $\delta/\Delta_c=-0.25$. However, if one zooms in locally, a local peak of $|t_{ba}|^2$ and a tiny one of $|t_{ab}|^2$ can be found at the position of the blue-dotted line. This illustrates that the first-order sideband (probe field) is magnified hundreds of times compared with the original input probe field and mainly transmitted at $\delta/\Delta_c=-0.3$ in case $1$. Whereas, in case $2$, magnification and transmission of the first-order sideband mainly occurs at $\delta/\Delta_c=-0.25$. From another point of view, the transmissions of the first-order sideband in two cases are robustly nonreciprocal near $\delta/\Delta_c=-0.3$. Continue to enhance the driving power to $30\,\,m$W in Fig.~\ref{fig:4}(c), we find $|t_{ba}|^2$ with two peaks is always smaller than $2$; while $|t_{ab}|^2$ reaches its maximum value $15.5$ at $\delta/\Delta_c=-0.31$, with the other invisible peak at the blue-dotted line. The results illustrate that the first-order sideband is magnified less than twice in case $1$, but more than ten times in case $2$; and the strong transmission nonreciprocity exists near $\delta/\Delta_c=-0.3$.
From above numerical analysis, we find the transmission nonreciprocity of the first-order sideband is effectively modulated by the driving power acted on the magnon mode. This stems from the fact that \textit{the increased driving power excites more magnon polarons
\begin{figure}[tbh]
\centerline{\includegraphics[width=8.5cm]{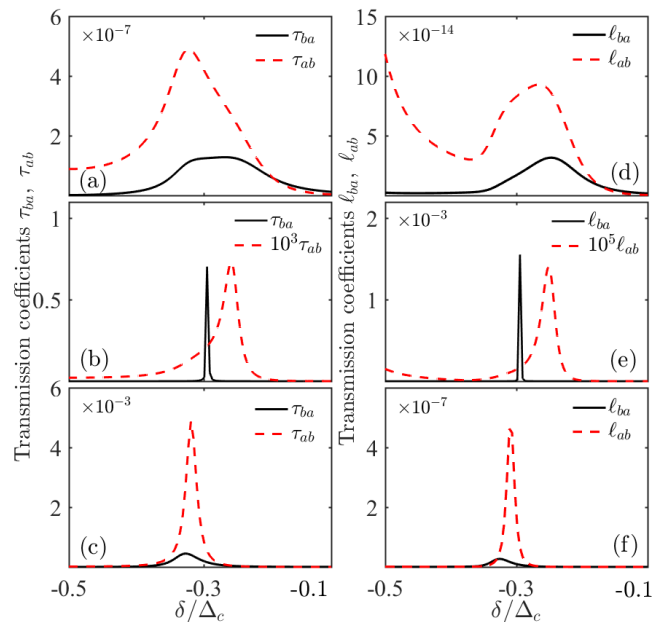}}
\caption{(Color online) Transmission coefficients $\tau_{ba}$, $\tau_{ab}$ and $\ell_{ba}$, $\ell_{ab}$ of the second- and third-order sidebands vs frequency detuning $\delta/\Delta_c$. In order to clearly observe the transmission coefficients, the values of $\tau_{ab}$ in (b) and $\ell_{ab}$ in (e) have been respectively magnified by three and five orders of magnitude. Here $P=0$ in panels (a) and (d), $P=20\,\,m$W in panels (b) and (e), and $P=30\,\,m$W in panels (c) and (f). Other parameters are same as that in Fig.~\ref{fig:2}.}
\label{fig:5}
\end{figure}
and simultaneously enhances the effective Kerr nonlinearity of the magnon}, \textit{which eventually strengthens the nonlinear asymmetric responses} of the system in two cases.
In addition, it needs to be emphasized that peaks on each curve should have been resonant with the effective supermodes made up of two microwave cavity modes. Frequencies of the supermodes shift $\pm\sqrt{J^2- (\frac{\kappa_b-\kappa_a}{2})^2}$ (equal to $\pm0.3\Delta_c$) from the original cavity frequency $\omega_c$. This is originated from the strong cavity-cavity coupling rate, i.e., $J > \kappa_a,\kappa_b$ in the system. Practically, the actural resonant frequency exhibit shifts due to the joint interference of the strong cavity-magnon coherent interaction ($g >J$) and the magnon Kerr nonlinearity~\cite{Kons2015}.

Figure~{\ref{fig:5}} shows transmission coefficients $\tau_{ba}$, $\tau_{ab}$ and
$\ell_{ba}$, $\ell_{ab}$ of the second- and the third-order sidebands versus frequency detuning $\delta/\Delta_c$ with different resonant driving powers.
When the driving field is absent in Figs.~\ref{fig:5}(a) and \ref{fig:5}(d), although the transmission coefficients are very tiny, the transmission nonreciprocity is very obvious in the regime $\delta/\Delta_c \in[-0.5\,\,-0.2]$.
Once the driving field is considered, both the transmission coefficients and nonreciprocity will be further enhanced.
For instance, $P=20\,\,m$W in Figs.~\ref{fig:5}(b) and \ref{fig:5}(e), $\tau_{ba}$ and $\ell_{ba}$ reach respective unique maximum $0.7$ and $1.5\times10^{-3}$ at $\delta/\Delta_c=-0.3$. While $\tau_{ab}$ and $\ell_{ab}$ keep much smaller maximums. Continue to increase the driving power to $30\,\,m$W in Fig.~\ref{fig:5}(c) and \ref{fig:5}(f). Both $\tau_{ba}$ and $\ell_{ba}$ have very tiny maximums. While $\tau_{ab}$ and $\ell_{ab}$ have soared to their peaks $4.8\times10^{-3}$ and $4.6\times10^{-7}$ at the point $\delta/\Delta_c=-0.31$. This illustrates that the resonant driving field can not only effectively manipulate the transmission nonreciprocity of the first-order sideband, but also of the second- and the third-order sidebands. What's more, the transmission nonreciprocity of the high-order sidebands have a cooperative and consistent characteristic when facing different driving powers. Besides being applied in high performance nonreciprocal directional-switching
\begin{figure}[htb]
\centerline{\includegraphics[width=8.5cm]{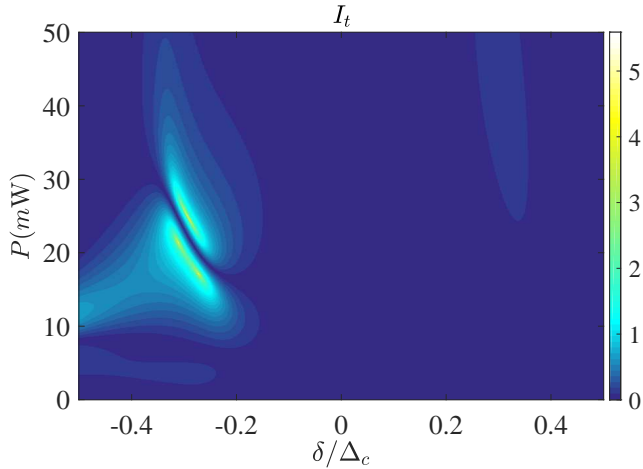}}
\caption{(Color online) The isolation ratio $I_t$ of the first-order sideband vs the resonant driving power $P$ and frequency detuning $\delta/\Delta_c$. Parameters are the same as that in Fig.~\ref{fig:2}.}
\label{fig:6}
\end{figure}
isolator~\cite{Assanto2008,Khanikaev2020} and diode~\cite{Gong2011,Zhu2013,Qi2012},
such a series of equally spaced high-order sidebands also have potential applications in frequency comb-like precision measurement through simply operating the driving power.

Then, we quantitatively describe the degree of transmission nonreciprocity for each sideband by isolation ratio. Figure~\ref{fig:6} intuitively displays isolation ratio $I_t$ varies with the resonant driving power $P$ and frequency detuning $\delta/\Delta_c$. As the driving power increases, one bright region, composed of an upper and a lower parts, appears at the center of $\delta/\Delta_c=-0.3$. In this region the optimal $I_t$ is $5.6$\,\,dB obtained with $P=21.8\,\,m$W. This can also be explained by the result in Fig.~\ref{fig:4}, i.e., the prominent large differences of transmission coefficients $|t_{ba}|^2$ and $|t_{ab}|^2$ between the left resonant peaks near $\delta/\Delta_c=-0.3$.

\begin{figure}[tbh]                                                                           \centerline{\includegraphics[width=9cm]{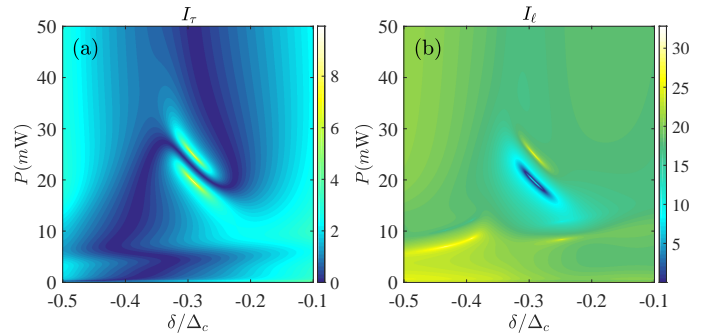}}
\caption{(Color online) isolation ratios (a) $I_{\tau}$ of the second-order sideband and (b) $I_{\ell}$ of the third-order sideband vs the resonant driving power $P$ and frequency detuning $\delta/\Delta_c$. Parameters are same as that in Fig.~\ref{fig:2}.}
\label{fig:7}
\end{figure}

For the second- and third-order sidebands, the corresponding isolation ratios $I_{\tau}$ and $I_{\ell}$ versus the driving power $P$ and frequency detuning $\delta/\Delta_c$ are given in Fig.~{\ref{fig:7}}(a) and {\ref{fig:7}}(b). In Fig.~{\ref{fig:7}}(a), the value of $I_{\tau}$ in most areas is lower than $3.5$\,\,dB,
except two bright areas centered at $\delta/\Delta_c=-0.3$, where $I_{\tau}$ reaches its maximum $9.9$\,\,dB.
While the distribution of $I_{\ell}$ is quite different. As presented in Fig.~{\ref{fig:7}}(b), except a small dark area,
the values of $I_{\ell}$ in other areas are higher than $10$\,\,dB. Especially in the yellow area centered at $\delta/\Delta_c=-0.4$ with $P\in[0\,\,10]\,m$W and the other one centered at $\delta/\Delta_c=-0.3$ with $P\in[22\,\,28]\,m$W, $I_{\ell}$ can be higher than $30$\,\,dB. By an overall observing, what $I_{\ell}$ and $I_{\tau}$ have in common is that the optimal value is distributed near $\delta/\Delta_c=-0.3$, this is consistent with the result in Fig.~{\ref{fig:5}}.

Taken together Fig.~{\ref{fig:6}} and Fig.~{\ref{fig:7}}, we find that the isolation ratios $I_{t}$, $I_{\tau}$ and $I_{\ell}$ show an overall increasing trend with the increase of the sideband order. Specifically, optimal $I_{t}$, $I_{\tau}$ and $I_{\ell}$ are enhanced in turn from $5.6$\,\,dB, to $9.9$\,\,dB and lastly $30$\,\,dB near detuning $\delta/\Delta_c=-0.3$. The reason is rooted from the strengthened effective Kerr nonlinearity in each sideband. Besides the steady-state amplitude of the magnon mode, the first-order sideband is also associated with the first-order perturbation terms (in equations~(\ref{e9})) of the magnon mode; the second-order sideband is simultaneously related to the first- and the
second-order perturbation terms of the magnon mode (in equations.~(\ref{e13})); and the third-order sideband is totally related to the first-, second- and third-order perturbation terms (in equations~(\ref{e18})) of the magnon mode. Except for the steady-state amplitude of the magnon, the strong transmission nonreciprocity of each sideband originates from the aggregate nonlinear effects in each perturbation term. Therefor, the higher the order of the sideband, the stronger nonlinear asymmetry of the transmission, it lastly leads to much higher isolation ratio.
\begin{figure}[tbh]
\centerline{\includegraphics[width=8.5cm]{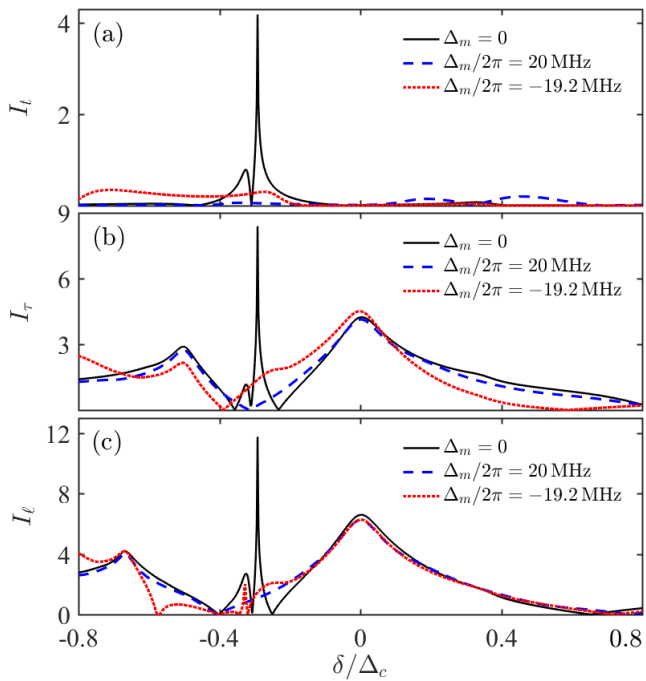}}
\caption{(Color online) The isolation ratios (a) $I_{t}$, (b) $I_{\tau}$ and (c) $I_{\ell}$ vs $\delta/\Delta_c$ with different detunings $\Delta_m$ between the magnon mode and the correspong driving field. Here $P=25\,\,m$W, and the other parameters are the same as in Fig.\,2.}
\label{fig:8}
\end{figure}

\subsection{Dependence of the nonreciprocal transmissions for high-order sidebands on frequency detuning $\Delta_m$}
In this section, we research influence of frequency detuning $\Delta_m$ on the nonreciprocal transmission of the high-order sidebands. In Fig.~{\ref{fig:8}, Isolation ratios $I_t,~I_{\tau}$ and $I_{\ell}$ vs $\delta/\Delta_c$ with different detunings $\Delta_m$ are intuitively given}.
When magnon mode is resonant with the driving field i.e., $\Delta_m=0$, isolation ratios $I_t,~I_{\tau}$ and $I_{\ell}$ increase in turn. Especially in the optimal detuning regime (centered at the resonant frequency of one effective supermode of the microwave cavities~\cite{Kons2015}), $I_t$, $I_{\tau}$ and $I_{\ell}$ reach their maximum values. The same evolution trend is also suitable for non-resonant situations, i.e., $\Delta_m/2\pi=-19.2$\,\,MHz, $20$\,\,MHz, except for the reduced optimal isolation ratios and the changed optimal detuning regimes. It is because \textit{excitation of the magnetic polarons is suppressed in the non-resonant situations and results in much weaker effective Kerr nonlinearity}. This lastly induces much weaker transmission nonreciprocity and shifts the optimal detuning regime.

\begin{figure}[tbh]
\centerline{\includegraphics[width=8.5cm]{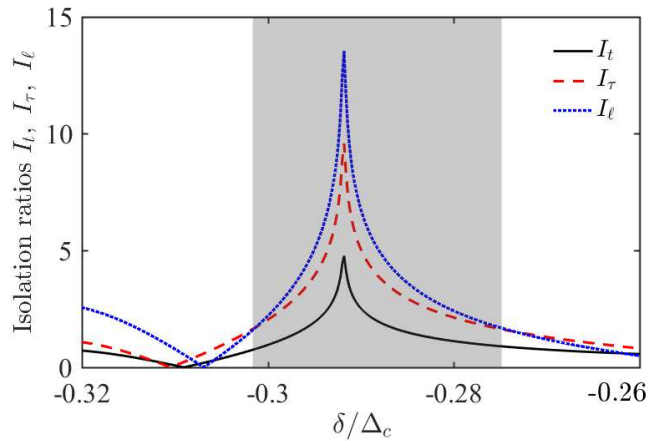}}
\caption{(Color online) The isolation ratios $I_{t}$, $I_{\tau}$ and (c) $I_{\ell}$ vs $\delta/\Delta_c$ when the magnon mode is resonant with the driving field, i.e., $\Delta_m=0$. Here $P=25\,\,m$W and the other parameters are the same as in Fig.\,2.}
\label{fig:9}
\end{figure}

For actual operation to obtain strong nonreciprocity of the high-order sidebands, the bandwidth of the optimal detuning regime is often more concerned. In the resonant situation, we centralize $I_t$, $I_{\eta}$ and $I_{\ell}$ in Fig.\,\,~{\ref{fig:9}}. It can be obviously seen $I_t$, $I_{\eta}$ and $I_{\ell}$
increase in turn and reach their optimal values in the gray shaded area. This means an overall controlling of the strong nonreciprocity for high-order sidebands can be realized in the optimal detuning regime ranging from $\delta/\Delta_c=-0.301 $ to $\delta/\Delta_c=-0.273$. The bandwidth of the optimal detuning regime is seven megahertz with $\Delta_c=2\pi\times40$\,\,MHz in the system.
It is achievable for experimental operation to capture strong transmission nonreciprocity of the high-order sidebands simultaneously in such a bandwidth.

\section{Conclusion}
In summary, we have theoretically provided a method to realize an overall controlling of the strong transmission nonreciprocity of the first-, second- and third-order sidebands in a cavity magnonics system. It is consist of two coupled microwave cavities and one YIG (or another suitable magnonic material) sphere. Our approach utilizes the self-Kerr nonlinearity of magnon in the YIG sphere. We have shown that strong transmission nonreciprocity of the high-order sidebands can be achieved by respectively regulating the driving power on the magnon mode and the frequency detuning between the magnon mode and the driving field. We also showed that the higher the order of the sideband, the stronger is the nonreciprocity marked by the enhanced isolation ratio in the optimal detuning regime. We have illustrated that the bandwidth of the optimal detuning regime can reach several megahertz when magnon mode is resonant drived. This implies that it is experimentally feasible to capture robust transmission nonreciprocity of the high-order sidebands simultaneously.
This study provides a promising route to realize strong nonreciprocity of the high-order sidebands at microwave frequencies and has potential applications in frequency comb-like precision measurement.
\begin{acknowledgements}
We acknowledge professor Xin-You L\"{u} for his valuable suggestions on our work. And this work was supported by the National Natural Science Foundation of China (NSFC) under Grants 12005078, 11675124 and 11704295.
\end{acknowledgements}

\end{document}